\documentclass{article}
\usepackage{spconf,amsmath,epsfig}
\usepackage{booktabs} 
\usepackage{threeparttable}
\usepackage{epstopdf}
\usepackage{graphicx}
\usepackage{multicol}
\usepackage{multirow}
\usepackage[tight,footnotesize]{subfigure}
\usepackage{algorithm}
\usepackage{algpseudocode}
\usepackage{amsmath}
\usepackage{graphicx}
\usepackage{subfigure}
\usepackage{cite}
\usepackage{balance}

\begin{document}
\title{Reinforcing Short-Length Hashing}
\name{Xingbo Liu$^{1}$, Xiushan Nie$^{2*}$, Qi Dai$^{3,}$,
Yupan Huang$^{4,}$,Yilong Yin$^{3}$   \thanks{* corresponding author}}
\address{$^1$School of Software, Shandong University, Jinan, P.R. China\\
$^2$School of Computer Science and Technology, Shandong Jianzhu University, Jinan, P.R. China\\
$^3$Microsoft Research Asia\\
$^4$School of Computer Science and Technology, Sun Yat-sen University,   Guangzhou, P.R. China\\
sclxb@mail.sdu.edu.cn, niexsh@sdufe.edu.cn, ylyin@sdu.edu.cn 
}

\maketitle

\begin{abstract}
Due to the compelling efficiency in  retrieval and storage, similarity-preserving hashing has been widely applied to approximate nearest neighbor search in large-scale image retrieval. However, existing methods have poor performance in retrieval using an extremely short-length  hash code due to weak ability of classification and poor distribution of hash bit. To address this issue, in this study, we propose a novel reinforcing short-length hashing (RSLH). In this proposed RSLH, mutual reconstruction between the hash representation and semantic labels is performed to preserve the semantic information. Furthermore, to enhance the accuracy of  hash representation, a pairwise similarity matrix is designed to make a balance between accuracy and training expenditure on memory. In addition, a parameter boosting strategy is integrated to reinforce the precision with hash bits fusion. Extensive experiments  on three large-scale image benchmarks demonstrate the superior performance of RSLH under various short-length hashing scenarios.
\end{abstract}

\section{Introduction}
In the era of big data, the approximate nearest neighbor (ANN) search that finds  ANNs of a query sample within a large database has become ubiquitous in numerous applications, such as image and video retrieval \cite{hao2017unsupervised} \cite{wang2018survey} . As a hot topic in information retrieval, hashing can provide an advantageous solution to ANN search based on its remarkable efficiency in both storage cost and query speed.

Hashing encodes high-dimension media data into a string of complex binary codes and preserves the similarity of original data at the same time. In contrast to other distance calculations \cite{wang2018survey}, the distance calculations in hashing utilize Hamming distance which can be implemented on hardware with bit-wise XOR operations, to provide higher efficiency.
Aiming at generating hash codes under the assistance of original data, learning-based hashing can provide advanced retrieval performance in ANN search. And the existing learning-based hashing methods can be roughly divided into two main categories: the unsupervised \cite{gong2011iterative}, \cite{Ding2014Collective} and the supervised  \cite{liu2012supervised} \cite{lin2015supervised} \cite{shen2015supervised}, \cite{gui2016supervised}, \cite{liu2016natural},  \cite{lai2018tip}, \cite{liu2019SSLH}, \cite{liu2019SDHMLR}.  In general, supervised hashing methods outperform  the unsupervised  dramatically by adopting semantic label information.

In learning-based hashing, one of the primary purposes is to obtain more compact and shorter hash codes with high precision. However, the retrieval accuracy will degrade dramatically if the length of hash code is extremely short. Therefore, how to learn a short hash code with higher accuracy is a challenge in hash learning field. In general, given a dataset with $c$ categories, the length $L$ of the hash code should be greater than $log_{2}(c)$; otherwise, the hash codes cannot distinguish the samples. In \cite{liu2019SSLH},  \emph{short-length} is defined as the integer length which is slightly greater than $log_{2}(c)$. For example, in a dataset which  has 10 categories, the length of four bits can be considered as a short length. Short-length hash codes can reduce the storage cost and computation complexity, thus accelerate the retrieval speed \cite{Luo2018Collaborative}. 

Generally, retrieving with short-length hash codes usually leads to poor performance due to the following reasons: 1) \emph{Weak ability of  classification}: Classification ability is the foremost in information retrieval, but hash codes with short-length suffer from  poor classification ability. Therefore, enhancing the classification ability is vital for short-length hashing. And 2) \emph{bad distribution of  hash bit}: A bad distribution of hash bit means the uncorrelation and balance constraints are kept badly, which will result in trivial solutions during hash learning.

In order to address the aforementioned issues, we propose a novel discrete hashing method, termed reinforcing short-length hashing (RSLH). In the proposed RSLH,  mutual regression between the hash codes and semantic labels is performed for enhancing  the  ability of classification. Moreover, to promote the accuracy of the hash representation, a pairwise similarity matrix is designed to make a balance between accuracy and training expenditure on memory. In addition, a parameter boosting strategy is integrated into hash learning for avoiding suboptimal solutions. 

The main contributions of this study are summarized as follows:
\begin{itemize}
\item We propose a supervised discrete  method for short-length hash learning. In this method, mutual regression, semantic pairwise similarity and relaxed strategy are seamlessly integrated for reinforcing the short-length hash learning.

\item A model boosting strategy is designed to improve the performance of short-length hash learning based on the uncorrelation and balance constraints.

\item Extensive experiments  on three large-scale datasets demonstrate that the proposed method  performs well under various short-length hashing scenarios.
\end{itemize}

\section{Proposed Method}

\subsection{Formulation}
Assume there is a training set consisting of $n$ instances, \emph{i.e.}, ${\bf{A}} = \{ {\bf{a}}_i\} _{i = 1}^n$, where each instance can be represented by a $m$-dimensional feature. Moreover, a class label matrix, ${\bf{Y}} = \{ {{\bf{y}}_i}\} _{i = 1}^n$, is available, with ${{\bf{y}}_i= \{ {{{y}}_{ij}}} \} \in {\{-1,1\}}^{c}$ being the label vector of the $i$-th instance, where $c$ is the number of categories. If the $i$-th instance belongs to the $j$-th category, ${y_{ij}} = 1$, and $-1$ otherwise. The hash matrix is defined as ${\bf{H}}=\{{\bf{h}}_{i}\}_{i = 1}^n$. ${\rm{||}}{\bf{H}}||$ and ${{\bf{H}}^T}$ mean the $\ell_{2}$-norm and transpose of matrix, $\bf{H}$, respectively.

In this study, a radial basis function is adopted to remedy the loss from feature information. The kernel trick makes sense by capturing the local structure of the data and  the feature dimension with nonlinear projections. Specifically, we first randomly select $d$ anchor points, ${\{{\bf{p}}_i\}_{i=1}^{d}}$, from the training set, then transform each training sample into a new representation by
\begin{equation}
 \label{solution-kernel}
\varphi ({\bf{a}})=exp(-\left \| {\bf{a}}-{\bf{p}}_i \right \|^{2}/2\sigma^{2} )_{i=1}^{d}.
\end{equation}
This process can be calculated ahead; $\varphi({\bf{A}})$ is  represented by $\bf{X}$ in the following sections for conciseness.

Given the hash code matrix ${\bf{H}}$ and label matrix ${\bf{Y}}$, a linear model is commonly-used to describe the correlation because of its efficiency. Typically, SDH \cite{shen2015supervised} adopts the projection from the hash matrix ${\bf{H}}$ to label matrix ${\bf{Y}}$, which can be formulated as
\begin{equation}
\label{Loss_B}
\min\limits_{{\bf {W}},{\bf {H}}}\left \| {\bf {Y}}-{\bf{W}}^{T}\bf{H} \right \|^{2},
  \quad \textup{s.t.}  \quad {\bf {H}} \in \left  \{ -1, +1 \right \}^{L\times n},
\end{equation}
where $L$ is the length of hash code. However, this strategy make the process of hashing learning  time-consuming and unstable in a way. To solve this problem, FSDH \cite{gui2018fast} attempts to learn a projection from label matrix ${\bf{Y}}$ to hash matrix ${\bf{H}}$, and it can be formulated as 
\begin{equation}
\label{Loss_H}
\min\limits_{{\bf {M}},{\bf {H}}}\left \| {\bf {H}}-{\bf{M}}^{T}{\bf{Y}} \right \|^{2},
  \quad \textup{s.t.}  \quad {\bf {H}} \in \left \{ -1, +1 \right \}^{L\times n}.
\end{equation}
In this study, we combine these two strategies to enhance the classification ability, which is formulated  as
\begin{equation}
\begin{split}
\label{Loss_HH}
&\min\limits_{{\bf {W}},{\bf {M}},{\bf {H}}}\left \| {\bf {Y}}-{\bf{W}}^{T}\bf{H} \right \|^{2}+ \alpha \left \| {\bf {H}}-{\bf{M}}^{T}{\bf{Y}} \right \|^{2},\\
&\quad \textup{s.t.}  \quad {\bf {H}} \in \left \{ -1, +1 \right \}^{L\times n}.
\end{split}
\end{equation}

Both the hash code and the class label can be considered as kinds of sample representations in Hamming space since both of them are binary. Therefore, the mutual regression between hash codes and class labels can be formulated as a linear auto-encoder process. Inspired by the study in \cite{liu2019SDHMLR}, we use the same projection for the regression loss between label matrix $\bf{Y}$ and hash matrix matrix multiplication $\bf{H}$ (i.e., ${\bf{W}}={\bf{M}}^T$ in Eq. (\ref{Loss_HH}), the transpose is used for matrix multiplication), and the similarity semantic similarity can be well preserved by using the same projection matrix between label matrix and hash matrix.

In addition, bit uncorrelation is an important constraint in  hash learning, which is formulated as 
\begin{equation}
 {\bf{H}}{{\bf{H}}^T} = n{\bf{I}}.
\end{equation}
Violating this constraint  leads to poor distribution of  hash bit, which has a great impact on short-length hashing. However, it is not suitable to directly integrate the uncorrelation constraint into discrete hashing method due to the following reasons: 1) discrete optimization is intractable since the binary quadratic programming is time-consuming and complicated; and 2) the hyperparameter is set empirically and difficult to influence the performance. To tackle this issue, we restrain ${\bf{H}}$ in Eq.(\ref{Loss_B}) to be  a real-valued orthogonal matrix ${\bf{B}}$, \emph{i.e.} ${\bf{B}}{\bf{B}}^{T}$ = ${\bf{I}}$. 


Furthermore, we assume $\bf{P}$ is a projection between  kernelized feature and real-valued orthogonal representation ${\bf{B}}$. In order to prevent over-fitting and improve the stability of regression \cite{hoerl1970ridge}, this study adopts the $\ell_{2}$-norm regularization for $\bf{P}$, and can be formulated as 
\begin{equation}
\label{Loss_P}
\min\limits_{{\bf{P}}}\left\|{\bf{B}}-{\bf{P}}^{T}{\bf{X}} \right\|^{2}+\lambda\left \|{\bf{P}}\right \|^{2},
\end{equation}
where $\lambda$ is a regularization parameter.

In short, the utilization of semantic label supervision can be formulated  as 
\begin{equation}
\label{Loss_Y}
\begin{split}
\min\limits_{{\bf {W}},{\bf {B}},{\bf {H}}}
&\left \|{\bf {Y}}-{\bf{W}}^{T}{\bf {B}}\right \|^{2}+ \alpha \left \|{\bf {H}}-{\bf{W}}{\bf {Y}}\right \|^{2}+ \beta \left \|{\bf {H}}-{\bf {B}}\right \|^{2}\\
&+\mu \left\|{\bf{B}}-{\bf{P}}^{T}{\bf{X}} \right\|^{2}+\lambda \left \|{\bf{P}}\right \|^{2}, \\
&\textup{s.t.}  \quad {\bf {B}}{\bf {B}}^{T} = {\bf{I}}, {\bf {H}}\in \{-1,+1\}^{L\times n}.
\end{split}
\end{equation}

It can be seen that only the hash matrix $\bf{H}$ in Eq. (\ref{Loss_B}) is relaxed, and the advantages are three-folds: 1) the discrete optimization in Eq. (\ref{Loss_B}) can be bypassed tactfully, making the optimization of ${\bf{H}}$ easier and faster; 2) the problem of solving ${\bf{W}}$ can be transformed form Sylvester equation \cite{Golub2009Matrix} to least square regression, saving  time for training. And 3) ${\bf{H}}$ in Eq. (\ref{Loss_H}) can be optimized discretely without quantization, making the proposed method more precise.
Details are showed in Section 2.3.

Furthermore, to capture more relations among samples and  generate more similarity-preserving hash representations, pairwise similarity matrix ${\bf{S}}$ is embedded into the Hamming space. Unlike previous study,  a novel asymmetric strategy is proposed in this study. Specifically, real-valued orthogonal representation ${\bf{B}}$ and binary code ${\bf{H}}$ are elaborated to preserve the pairwise similarity. This process can be formalized as follows 
\begin{equation}
\label{Loss_S}
\min\limits_{{\bf {B}},{\bf {H}}}\left \|{\bf {B}}^{T}{\bf {H}}- {\bf {S}}\right \|^{2},
 \quad \textup{s.t.}  \quad {\bf {B}}{\bf {B}}^{T} = {\bf{I}}, {\bf {H}} \in \left \{ -1, +1 \right \}^{L\times n}.
\end{equation}

However, the pairwise similarity matrix ${\bf{S}}$ is of $n \times n$ size, making the training process much expensive on space. To tackle this problem, we prefer a predefined projection matrix ${\bf{R}}^{n \times k}$ ($k<n $) to decrease the expenditure, and the Eq. (\ref{Loss_S}) is formalized as follows 
\begin{equation}
\label{Loss_G}
\min\limits_{{\bf {B}},{\bf {H}}}\left \|{\bf {B}}^{T}{\bf {H}}{\bf{R}}- {\bf{SR}}\right \|^{2},
\textup{s.t.}  \quad {\bf {B}}{\bf {B}}^{T} = {\bf{I}}, {\bf {H}} \in \left \{ -1, +1 \right \}^{L\times n}.
\end{equation}

It can be proved that Eq. (\ref{Loss_S}) and Eq. (\ref{Loss_G}) are approximately equivalent to each other when ${\bf {R}}{\bf {R}}^{T} = {\bf{I}}^{n \times n}$. And the simple proof is presented as follows.
Problem in Eq.(\ref{Loss_S}) can be reformulated as
\begin{equation}
\label{Loss_SG_1}
\begin{split}
&\min\limits_{{\bf {B}},{\bf {H}}}\left \|{\bf {B}}^{T}{\bf {H}}- {\bf {S}}\right \|^{2}=\min\limits_{{\bf {B}},{\bf {H}}}\left \|{\bf {B}}^{T}{\bf {H}}\right \|^{2}+\left \|{\bf {S}}\right \|^{2}
-2Tr({\bf {H}}^{T}{\bf {B}}{\bf {S}})\\
&=\min\limits_{{\bf {B}},{\bf {H}}}Tr({\bf {H}}^{T}{\bf {B}}{\bf {B}}^{T}{\bf {H}}+{\bf {S}}^{T}{\bf {S}}-2{\bf {H}}^{T}{\bf {B}}{\bf {S}}).
\end{split}
\end{equation}

Problem in Eq.(\ref{Loss_G}) can be reformulated as
\begin{equation}
\label{Loss_SG_2}
\begin{split}
&\min\limits_{{\bf {B}},{\bf {H}}}\left \|{\bf {B}}^{T}{\bf {HR}}- {\bf {SR}}\right \|^{2}=\min\limits_{{\bf {B}},{\bf {H}}}\left \|{\bf {B}}^{T}{\bf {HR}}\right \|^{2}+\left \|{\bf {SR}}\right \|^{2}\\
&-2Tr({\bf {R}}^{T}{\bf {H}}^{T}{\bf {B}}{\bf {SR}})
=\min\limits_{{\bf {B}},{\bf {H}}}Tr({\bf {H}}^{T}{\bf {B}}{\bf {B}}^{T}{\bf {H}}{\bf {R}}{\bf {R}}^{T}\\
&+{\bf {S}}^{T}{\bf {S}}{\bf {R}}{\bf {R}}^{T}-2{\bf {H}}^{T}{\bf {B}}{\bf {S}}{\bf {R}}{\bf {R}}^{T}).
\end{split}
\end{equation}
Obviously, the Eq. (\ref{Loss_SG_1}) and Eq. (\ref{Loss_SG_2}) are equivalent to each other when ${\bf {R}}{\bf {R}}^{T} = {\bf{I}}^{n \times n}$. In this study, we define the  projection ${\bf {R}}$ as an orthogonal projection from pairwise similarity ${\bf{S}}$ to the feature matrix ${\bf{X}}$. This process is written as follows 
\begin{equation}
\label{Loss_R}
\min\limits_{{\bf {R}}}\left \|{\bf{SR}}-{\bf{X}}\right \|^{2},
 \quad \textup{s.t.}  \quad {\bf {R}}{\bf {R}}^{T} = {\bf{I}}^{d \times d}.
\end{equation}
This problem can be solved by singular value decomposition (SVD) on ${\bf {S}}{\bf {X}}^{T}$ (similar details can be seen in B-step of Section 2.3); we use $\bf{G}$ to represent ${\bf{SR}}$ in the following sections for conciseness. Different from previous work \cite{luo2018fast}, the utilization of pre-computed matrices $\bf{G}$ and $\bf{R}$ is more generalized. Furthermore, we also try to reconstruct the feature matrix via hash representations and thus decrease the information loss from original data. 

In summary, the objective function of the proposed method can be formulated as
\begin{equation}  
\begin{split}
\label{LossFunction}
&\min\limits_{{\bf {W}},{\bf {B}},{\bf {H}},{\bf {P}}}
\left \|{\bf {Y}}-{\bf{W}}^{T}{\bf {B}}\right \|^{2}+\alpha \left \|{\bf {H}}-{\bf {W}}{\bf{Y}} \right \|^{2} \\
&+\beta \left \|{\bf {H}}- {\bf{B}}\right \|^{2}+\gamma\left \|{\bf{B}}^{T}{\bf {H}}{\bf {R}}-{\bf {G}}\right \|^{2} \\
&+\mu \left \|{\bf {B}}-{\bf {P}}^{T} {\bf{X}} \right \|^{2} +\lambda\left \|{\bf{P}}\right \|^{2} \\
&\textup{s.t.} \quad {\bf {B}}{\bf {B}}^{T} = {\bf{I}}, {\bf {H}}\in \{-1,+1\}^{L\times n},
\end{split}
\end{equation}
where $\alpha$, $\beta$, $\gamma$ and $\mu$ are hyperparameters.

\subsection{Model Boosting}
For more performance enhancement of short-length hash learning, in this study, we propose a hash boosting strategy by considering uncorrelation and balance constraints  to learn more optimized  hash codes. On the strength of  this optimized hash code, we learn a new projection for out-of-sample extension.

Uncorrelation and bit balance are two important constraints in hash learning. The uncorrelation  constraint is formulated as Eq. (5). Here, we will briefly describe the bit balance  constraint.

Bit balance means that each bit has an approximately 50\% chance of being $+1$ or $-1$, which can be formulated as
${\bf{H1}} = 0$, 
where $\bf{1}$ is an $N$-dimensional all-ones vector.
Given a hash matrix ${\bf{H}} \in {\{-1,+1\}^{L \times N}}$, the balance degree of the $i_{th}$ bit for the samples can be defined as the absolute value of the sum of the $i_{th}$ row in the hash matrix. For example, if the vector $\{  - 1,{\kern 1pt} {\kern 1pt} 1,{\kern 1pt} {\kern 1pt} -1,{\kern 1pt} {\kern 1pt} -1\} $ is the $i_{th}$-row of the hash matrix, then the balance degree of the $i_{th}$ bit for the samples is $|-1+1-1-1|=2$. Obviously, the smaller balance degree indicates the better code balance, which means that it is more consistent with the balance constrain. When the hash bit is balanced, the entropy and information content reaches the maximum with small similarity loss, thereby demonstrating that the hash bit is superior~\cite{cao2018binary} \cite{liu2019moboost}. 

In this study, we propose a boosting strategy to obtain superior hash bits based on the uncorrelation and balance constrains during hash learning. The study in \cite{liu2019moboost} has proposed a boosting strategy called MoBoost. However, the MoBoost framework cannot handle with short-length hash codes effectively. The original MoBoost  framework destroys the uncorrelation constraint of hash bit. In this study, we propose a new strategy framework by uniting the uncorrelation
with balance constraint into one framework without hyperparameters. The process is described as follows.

Running the proposed method $T$ times, we can obtain $T$ hash matrices for the training samples. Then, we concatenate them in the column direction, and a new hash matrix with size $TL*n$ is acquired, where $L$ and $n$ are the short length of hash code and the number of samples, respectively.

The purpose of the proposed boosting strategy is to select $L$ rows from the concatenated matrix which have better bit balance and uncorrelation. Therefore, we use balance degree and clustering to achive this goal.
We take each row of the concatenated hash  matrix (i.e., each dimension of hash code) as a new instance, and then  perform the spectral clustering \cite{ng2002spectral} on these $TL$ new instances to get $L$ clusters. It is noteworthy that  the proposed  framework is sufficiently general to utilize other clustering methods. In this study, the spectral clustering is employed for
illustration. In each cluster, we first select the balance hash bit which has the smallest balance degree, and then concatenate the selected $L$ hash bit to get the final hash matrix of training set. Obviously, the obtained hash bits for the training set are not only balance but also uncorrelation because they come from different clusters.

Finally, based on the final hash matrix of training samples, we learn a linear projection between the features and the  hash codes for the out-of-sample extension. We set the value of $T$ to three in  the experiments.

\subsection{Optimization}
It is intractable to optimize Eq. (\ref{LossFunction}) directly since it is noncontinuous and nonconvex. In this study, we try to solve this nondifferentiable problem using an iterative framework with the following steps.

{\bf {W-Step}}: Learn the projection, $\bf{W}$, with the other variables fixed. The problem in Eq. (\ref{LossFunction}) becomes
\begin{equation}
\label{solution-W1}
\min\limits_{{\bf {W}}}
\left \|{\bf {Y}}-{\bf{W}}^{T}{\bf {B}}\right \|^{2}+\alpha \left \|{\bf {H}}-{\bf {W}}{\bf{Y}} \right \|^{2}.
\end{equation}
Eq. (\ref{solution-W1}) can be reformulated as
\begin{equation}
\label{solution-W2}
\min\limits_{{\bf {W}}}
\left \|{\bf {W}} \right \|^{2}-2Tr({\bf {Y}}^{T}{\bf{W}}^{T}{\bf {B}})+\alpha (\left \|{\bf {W}}{\bf{Y}} \right \|^{2}-2Tr({\bf {H}}^{T}{\bf {W}}{\bf{Y}})).
\end{equation}
Setting the derivative of Eq. (\ref{solution-W2}) \emph{w.r.t} $\bf{W}$ as zero yields
\begin{equation}
\label{solution-W}
{\bf {W}}=({\bf{B}}{\bf{Y}}^{T}+\alpha {\bf{H}}{\bf{Y}}^{T} )^{-1}(\alpha{\bf{Y}}{\bf{Y}}^{T}+{\bf{I}}).
\end{equation}

{\bf {B-Step}}: Learn the orthogonal real-valued representation, $\bf{B}$, with the other variables fixed. The problem in Eq. (\ref{LossFunction}) becomes
\begin{equation}  
\begin{split}
\label{solution-B1}
&\min\limits_{{\bf{B}}}
\left \|{\bf {Y}}-{\bf{W}}^{T}{\bf {B}}\right \|^{2}+\beta \left \|{\bf {H}}- {\bf{B}}\right \|^{2}+\gamma\left \|{\bf{B}}^{T}{\bf {H}}{\bf {R}}-{\bf {G}}\right \|^{2} \\
&+\mu \left \|{\bf {B}}-{\bf {P}}^{T}{\bf{X}} \right \|^{2}, \quad \textup{s.t.} \quad {\bf {B}}{\bf {B}}^{T} = {\bf{I}}.
\end{split}
\end{equation}
 Eq. (\ref{Solution-B2}) can be rewritten as
\begin{equation}
\label{Solution-B2}
\begin{split}
\min\limits_{{\bf {B}}}
&-2Tr({\bf {Y}}^{T}{\bf{W}}^{T}{\bf {B}})+\left \|{\bf{W}}^{T}{\bf {B}}\right \|^{2}+ \beta(-2Tr({\bf {H}}^{T}{\bf {B}})\\
&+\left \|{\bf {B}}\right \|^{2})
+\gamma(\left \|{\bf {B}}^{T}{\bf {HR}}\right \|^{2} -2Tr({\bf {R}}^{T}{\bf {H}}^{T}{\bf{B}}{\bf {G}})), \\
&+\mu(\left \|{\bf {B}}\right \|^{2}-2Tr({\bf {B}}^{T}{\bf {P}}^{T}{\bf {V}})) \\
&\textup{s.t.}  \quad {\bf {B}}{\bf {B}}^{T} = {\bf{I}}.
\end{split}
\end{equation} 
Since $\left \|{\bf{W}}^{T}{\bf {B}}\right \|^{2}=Tr({\bf{W}}{\bf{W}}^{T})$, $\left \|{\bf {B}}\right \|^{2}=Tr({\bf {B}}{\bf {B}}^{T})=L$ and $\left \|{\bf{B}}^{T}{\bf {HR}}\right \|^{2}=Tr({\bf{R}}^{T}{\bf{H}}{\bf{B}}{\bf{B}}^{T}{\bf{HR}})=L \ast n$, Eq. (\ref{Solution-B2}) can be reformulated as
\begin{equation}
\label{Solution-B3}
\max\limits_{{\bf {B}}} Tr({\bf{Q}}{\bf {B}}),
\quad \textup{s.t.}  \quad {\bf {B}}{\bf {B}}^{T} = {\bf{I}}.
\end{equation} 
where ${\bf{Q}}={\bf {Y}}^{T}{\bf{W}}^{T}+\beta {\bf{H}}^{T}+\gamma {\bf {G}}{\bf {R}}^{T}{\bf {H}}^{T}+\mu {\bf {X}}^{T}{\bf{P}}$.
This problem is a Procrustes problem with analytic solutions \cite{Xia2015Sparse}. First we perform
SVD ${\bf{Q}} = {\bf{U}} \sum {\bf{V}}^{T}$, where ${\bf{U}}$ is an $n \times n$  orthogonal matrix, $\sum$ is an $n \times L$ matrix and ${\bf{V}}$ is an $L \times L$ orthogonal matrix. 
Then the solution for ${\bf{B}}$ is 
\begin{equation}
\label{solution-B}
{\bf{B} = {\bf{V}}\hat{{\bf{U}}}}^{T},
\end{equation}
where $\hat{{\bf{U}}}$ contains first $L$ columns of ${\bf{U}}$.

{\bf {H-Step}}: Learn the binary code, ${\bf{H}}$, with the other variables fixed. The problem in Eq. (\ref{LossFunction}) becomes
\begin{equation}
 \label{solution-H1}
\begin{split}
&\min\limits_{{\bf {H}}}
\left \|{\bf {H}}-{\bf {W}}{\bf{Y}} \right \|^{2} +\gamma \left \|{\bf{B}}^{T}{\bf {H}}{\bf {R}}-{\bf {G}}\right \|^{2}+\beta \left \|{\bf {H}}- {\bf{B}}\right \|^{2} \\
&\textup{s.t.} \quad  {\bf {H}}\in \{-1,+1\}^{L\times n}.
\end{split}
\end{equation}
Since  $\left \|{\bf {H}}\right \|^{2}=\left \|{\bf{B}}^{T}{\bf {H}}{\bf {R}}\right \|^{2}=L \ast n$, Eq. (\ref{solution-H1}) can be reformulated as
\begin{equation}
 \label{solution-H2}
\begin{split}
&\min\limits_{{\bf {H}}} -Tr({\bf {H}}^{T}({\bf {W}}{\bf{Y}}+\beta {\bf {B}}+\gamma {\bf {B}}{\bf {G}}{\bf {R}}^{T}))\\
&\textup{s.t.} \quad {\bf{H}}\in \left \{ -1,+1 \right \}^{L\times n}.
\end{split}
\end{equation}
The analytic solution of ${\bf{H}}$ can be calculated as
\begin{equation}
\label{solution-H}
    {\bf{H}}=sgn({\bf {W}}{\bf{Y}}+\beta {\bf {B}}+\gamma {\bf {B}}{\bf {G}}{\bf {R}}^{T}),
\end{equation}
where $sgn(\cdot)$ is a sign function.

 
{\bf {P-Step}}: Learn the projection matrix, ${\bf{P}}$, while holding the other variables fixed. The problem in Eq. (\ref{LossFunction}) becomes
 \begin{equation}
\min\limits_{{\bf{P}}}\left\|{\bf{B}}-{\bf{P}}^{T}{\bf{X}} \right\|^{2}+\lambda \left\|{\bf{P}}\right\|^{2}.
\end{equation}
The closed-form solution of ${\bf{P}}$ is
 \begin{equation}
  \label{solution-P}
{\bf{P}}=({\bf{VV}}^{T}+\lambda {\bf{I}})^{-1}{\bf{V}}{\bf{B}}^{T}.
 \end{equation}

In conclusion, we try to solve the problem of nonconvex mixed integer optimization based on the above steps. 
Convergence is reached within a few iterations, which is demonstrated in the Experiments section. 


\begin{table*}[htp]
  \centering
  \fontsize{9}{11}\selectfont
  \begin{threeparttable}
  \begin{tabular}{c|c|c|c|c|c|c|c|c|c|c|c|c}
    \toprule
     \multirow {2}{*}{Method} &\multicolumn{4}{|c}{CALTECH-101}  &\multicolumn{4}{|c}{CIFAR-10} &\multicolumn{4}{|c}{ImageNet-100}\cr
    \cmidrule(lr){2-5} \cmidrule(lr){6-9}\cmidrule(lr){10-13} 
     &\!8 bits\!&\!10 bits\!&\!12 bits\!&\!14 bits  \!&4 bits\!&6 bits\!&\!8 bits\!&\!10 bits \!&8 bits\!&10 bits\!&\!12 bits\!&\!14 bits \cr
    \midrule
    SH&0.2043&0.2053&0.2450&0.2744&0.2564&0.2801&0.2772&0.2838&0.0316&0.0338&0.0357&0.0424\cr
    PCAITQ&0.0535&0.0715&0.0920&0.1001&0.1589&0.2117&0.2286&0.2658&0.0266&0.0269&0.0271&0.0276\cr
    PCARR&0.0999&0.1184&0.1138&0.1295&0.1696&0.2302&0.2796&0.2674&0.0328&0.0331&0.0331&0.0334\cr
    MFH&0.2030&0.2208&0.2395&0.2538&0.2434&0.2549&0.2561&0.2702&0.0314&0.0342&0.0368&0.0388\cr
    SDH&0.2110&0.2339&0.2503&0.2904&0.2554&0.3176&0.4179&0.4986&0.0345&0.0435&0.0398&0.0443\cr
    NSH&0.3516&0.3887&0.4129&0.4323&0.3449&0.4842&0.5395&0.5798&\underline{0.1065}&0.1290&0.1514&0.1752\cr
    FSSH&0.3502&0.3915&0.4287&\underline{0.4475}&0.4015&0.5176&0.5708&\underline{0.6088}&0.0926&0.1307&0.1545&0.1792\cr
    SSLH &\underline{0.3695}&\underline{0.4078}&\underline{0.4330}&0.4458&\underline{0.4356}&\underline{0.5215}&0.5722&0.6058&0.1015&\underline{0.1345}&0.1592&\underline{0.1813}\cr
    SDHMLR&0.3584&0.3859&0.4129&0.4116&0.3417&0.4848&\underline{0.5875}&0.6027&0.1032&0.1222&\underline{0.1627}&0.1736\cr
    \hline
    RSLH&\bf{0.4263}&\bf{0.4469}& \bf{0.4712}&\bf{0.4837}&\bf{0.4688 }& \bf{ 0.5934}&\bf{0.6163}& \bf{0.6473}& \bf{0.1358 }&\bf{0.1749}& \bf{0.2089 }&\bf{0.2330} \cr
 \midrule[0.8 pt]
    SH&0.0915&0.1130&0.1204&0.1254&0.1958&0.2177&0.2449&0.2670&0.0239&0.0289&0.0345&0.0501\cr
    PCA-ITQ&0.0405&0.0408&0.0418&0.0453&0.2268&0.2331&0.2359&0.2462&0.0129&0.0129&0.0130&0.0131\cr
    PCA-RR&0.0572&0.0700&0.0612&0.0745&0.2326&0.2143&0.2149&0.2191&0.0134&0.0162&0.0156&0.0177\cr
    MFH&0.1169&0.1257&0.1390&0.1568&0.1905&0.2136&0.2375&0.2573&0.0206&0.0253&0.0322&0.0397\cr
    SDH&0.2332&0.1622&0.2458&0.2682&0.2855&0.3494&0.4176&0.5221&0.0189&0.0196&0.0256&0.0285\cr
    NSH&0.2924&0.3140&0.3234&0.3480&0.3385&0.4045&0.4956&0.5540&0.0776&0.1131&0.1534&0.1964\cr
    FSSH&0.3137&\underline{0.3704}&\underline{0.4024}&\underline{0.4350}&0.3801&0.4626&0.5367&0.5892&0.0755&0.1173&0.1621&0.2013\cr
    SSLH&0.3051&0.3339&0.3530&0.3706&\underline{0.3877}&\underline{0.4772}&\underline{0.5515}&0.5869&\underline{0.0853}&\underline{0.1190}&\underline{0.1629}&\underline{0.2086}\cr
    SDHMLR&\underline{0.3298}&0.3628&0.3891&0.4211&0.3821&0.4463&0.5414&\underline{0.6028}&0.0686&0.0971&0.1424&0.1885\cr
    \hline
    RSLH &\bf{0.3807}&\bf{0.4242}&\bf{0.4560}&\bf{0.4897}&\bf{0.4132}&\bf{0.5196}&\bf{0.6002}&\bf{0.6370}&\bf{0.1121}&\bf{0.1524}&\bf{0.2063}&\bf{0.2207}\cr 
    \bottomrule
    \end{tabular}
        \caption{The top panel shows the performance in terms of mAP scores on three benchmark datasets.  The bottom panel shows the performance in terms of mAP@H$\leq$2 scores. The best results for mAP and mAP@H$\leq$2 scores are shown in bold. The second best results for mAP and mAP@H$\leq$2 scores are shown with underlines.}
   \end{threeparttable}
\end{table*}


\section{Experiments}
This section will demonstrate experimental settings and results. The experiments were performed on a computer with an Intel(R) Core(TM) i7-6700 CPU and 32-GB RAM. Extensive experiments were conducted on three large-scale image datasets to  verify the effectiveness of the proposed method.

\subsection{Datasets and Experimental Settings}
Three extensively-used image benchmarks were utilized in the experiments, including  CALTECH-101 
\cite{fei2007learning}, CIFAR-10 
\cite{krizhevsky2009learning} and ImageNet-100 
\cite{russakovsky2015ImageNet}.

For the CIFAR-10 and ImageNet-100 datasets, we used  CNN-F model \cite{chatfield2014return} to perform feature learning. For the CALTECH-101, each image was represented as a 512-dimension GIST feature.
Our method was performed ten runs and the performances were averaged for comparison. As the experimental parameters, we empirically set $\alpha=3$, $\beta=10^{-2}$, $\gamma=10^{-5}$,  $\mu=10^{-5}$ and $\lambda=10^{-6}$.

\subsection{Evaluation Metric}
To evaluate the proposed method, we used evaluation matrices called mean average precision (mAP) and mean average precision@Hamming radius $\leq$ 2 (mAP@H$\leq$2). mAP includes the mean of the average precision (AP) values obtained for the top retrieved samples. By restricting the Hamming radius $\leq$ 2,  hashing can retrieve any bucket in the hash table in $O(1)$ time complexity by table lookups, which enables the most efficient constant-time retrieval. 
Moreover, we adopted  precision score to evaluate the performance of the proposed RSLH and other methods. 

\begin{figure}[htb]
\centering
\subfigure[ CALTECH-101]{
\includegraphics[width=0.138\textwidth]{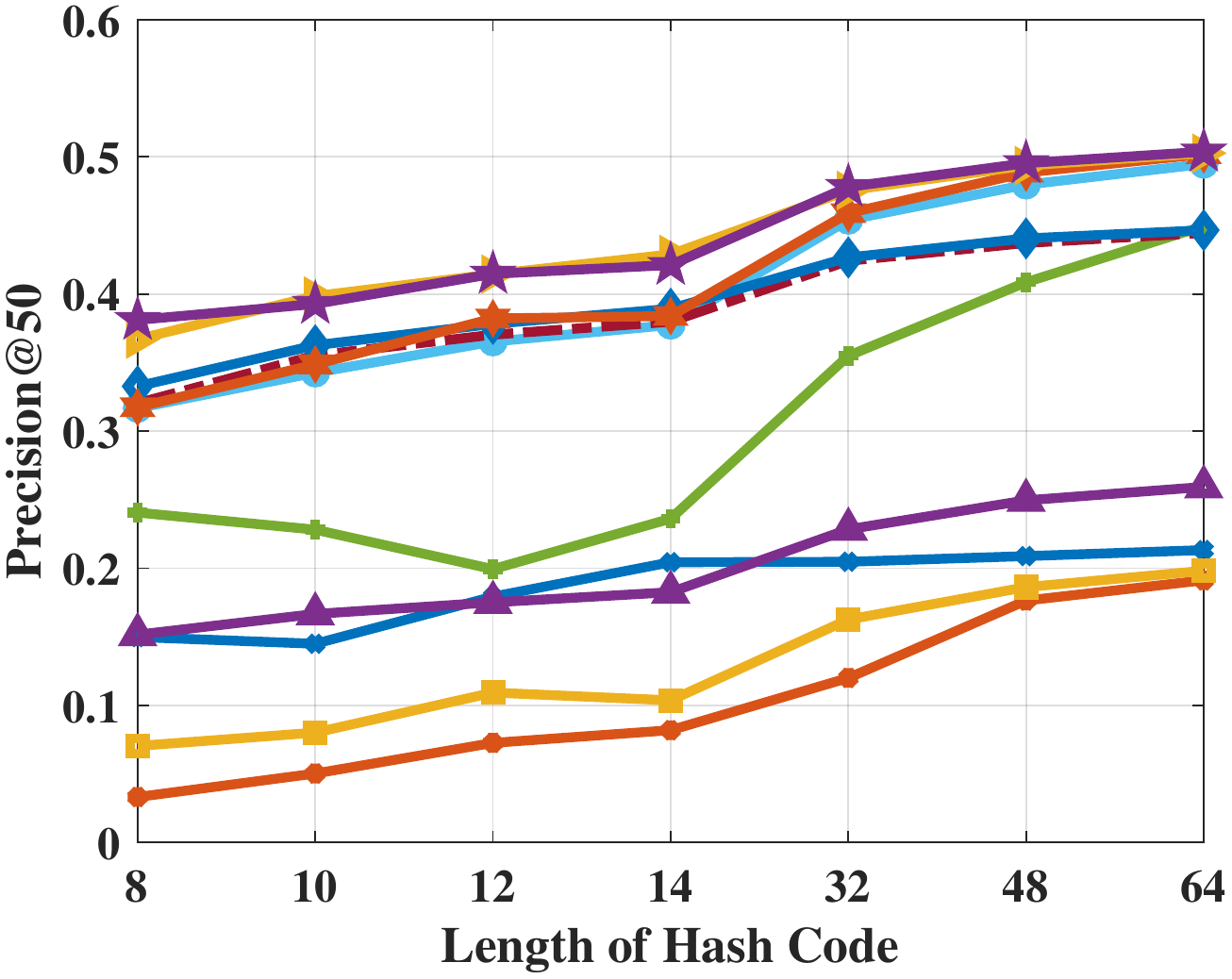}}
\subfigure[ CIFAR-10]{
\includegraphics[width=0.138\textwidth]{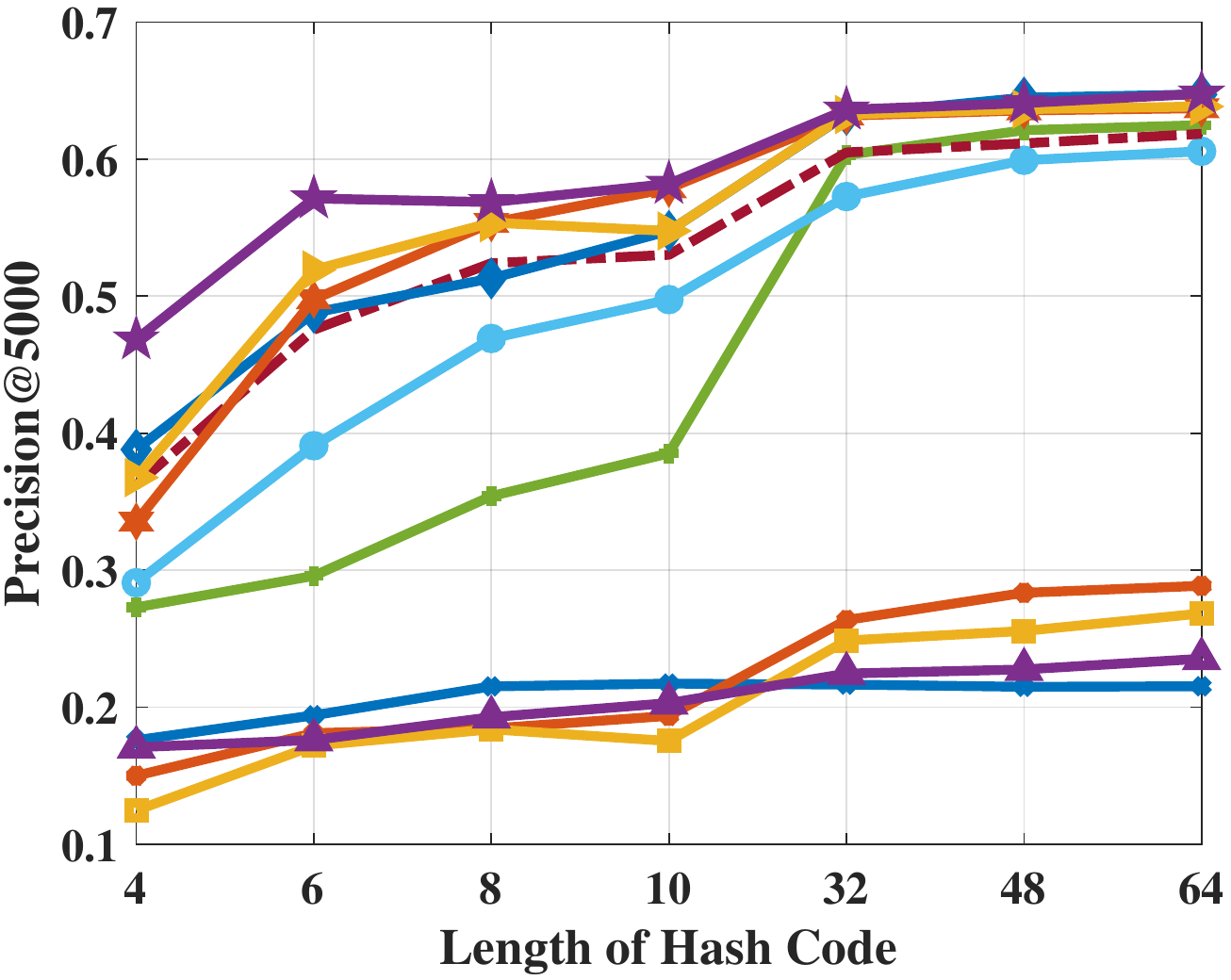}}
\subfigure[ImageNet-100]{
\includegraphics[width=0.178\textwidth]{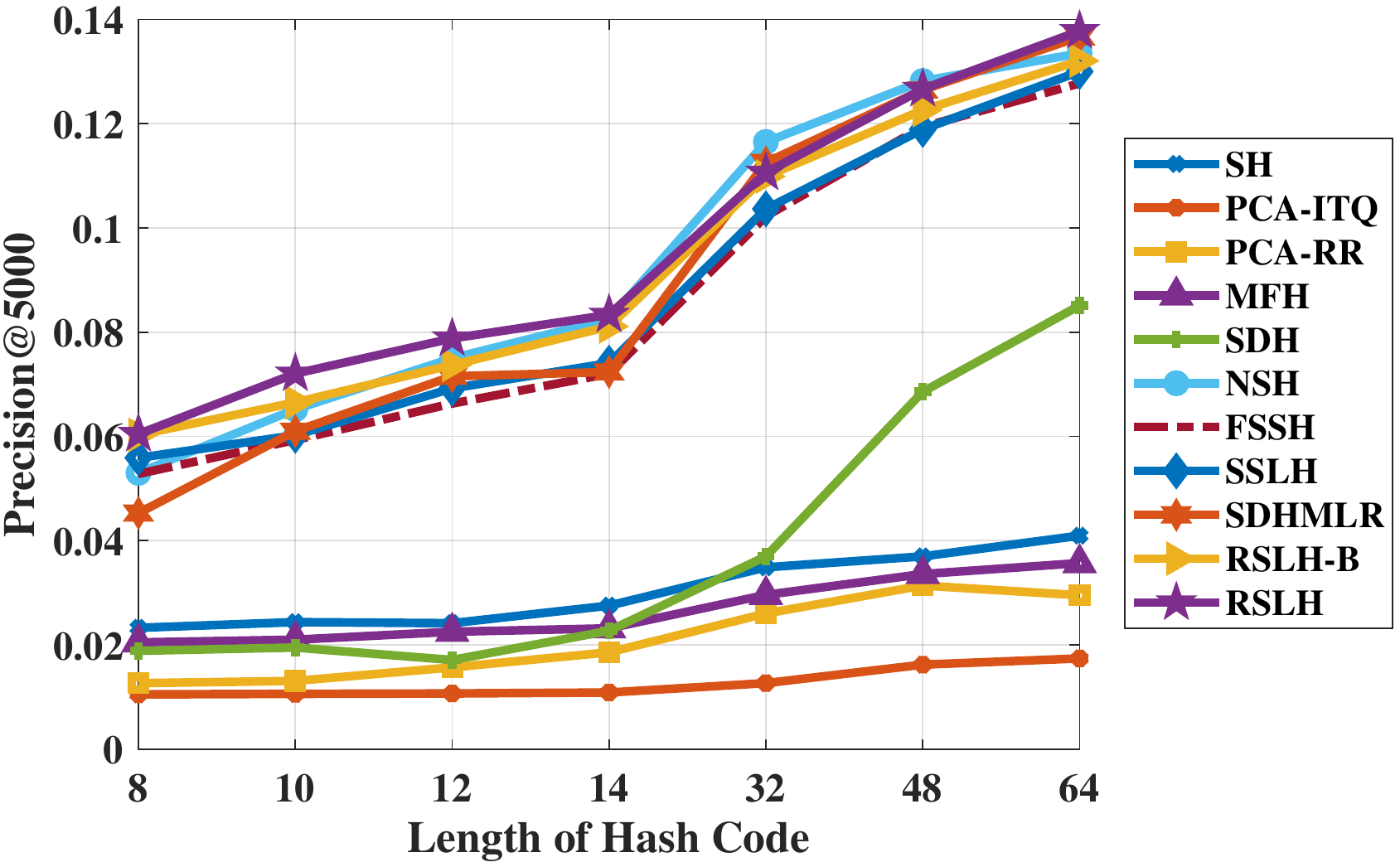}}

\subfigure[ CALTECH-101]{
\includegraphics[width=0.152\textwidth]{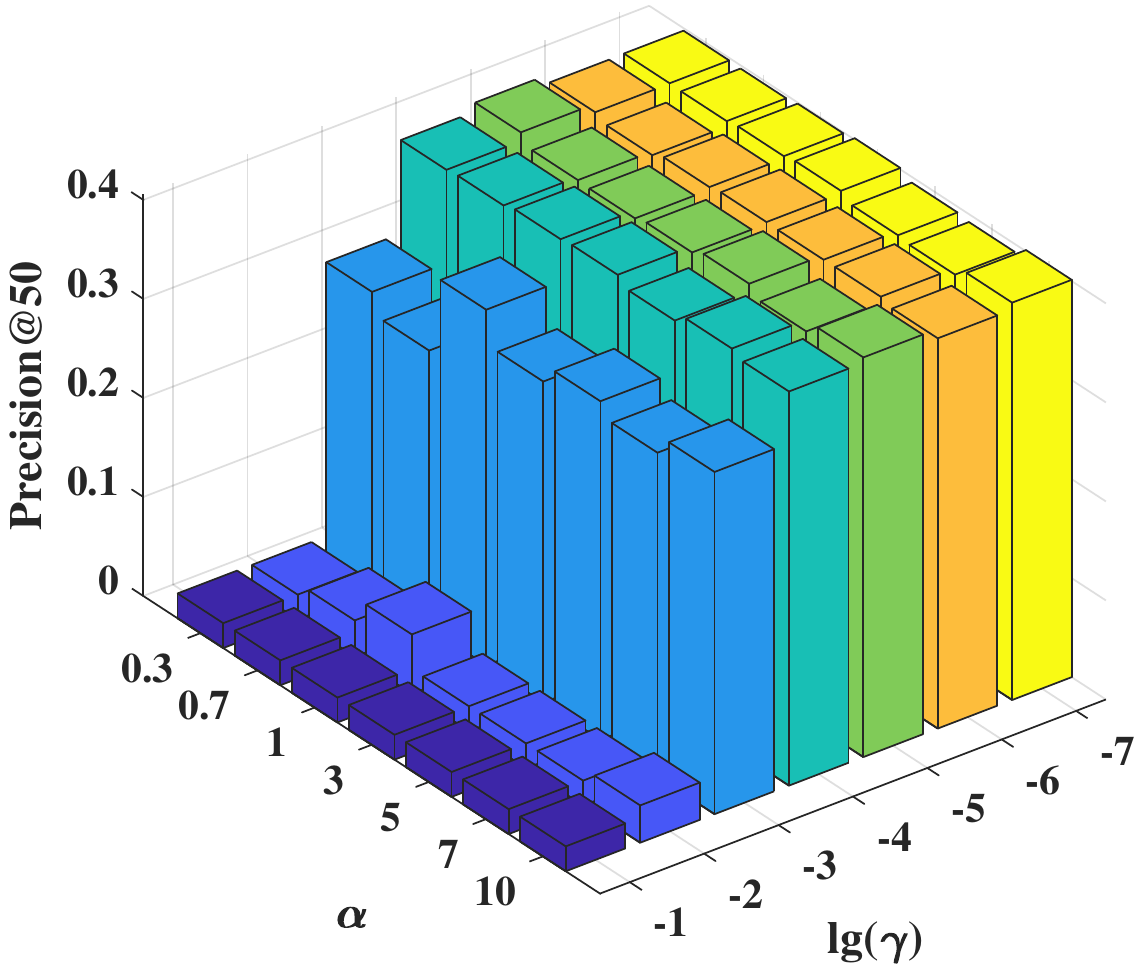}}
\subfigure[CIFAR-10]{
\includegraphics[width=0.152\textwidth]{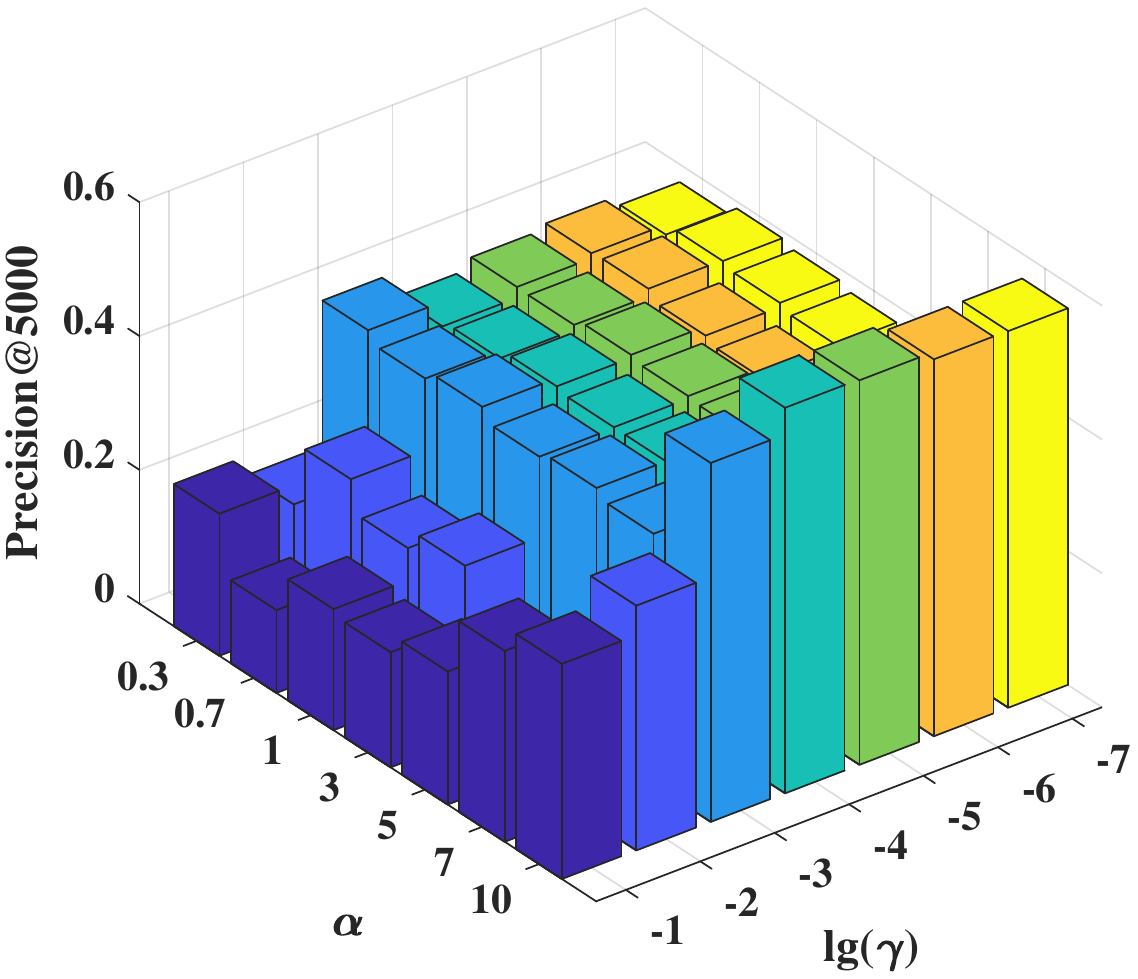}}
\subfigure[ImageNet-100]{
\includegraphics[width=0.152\textwidth]{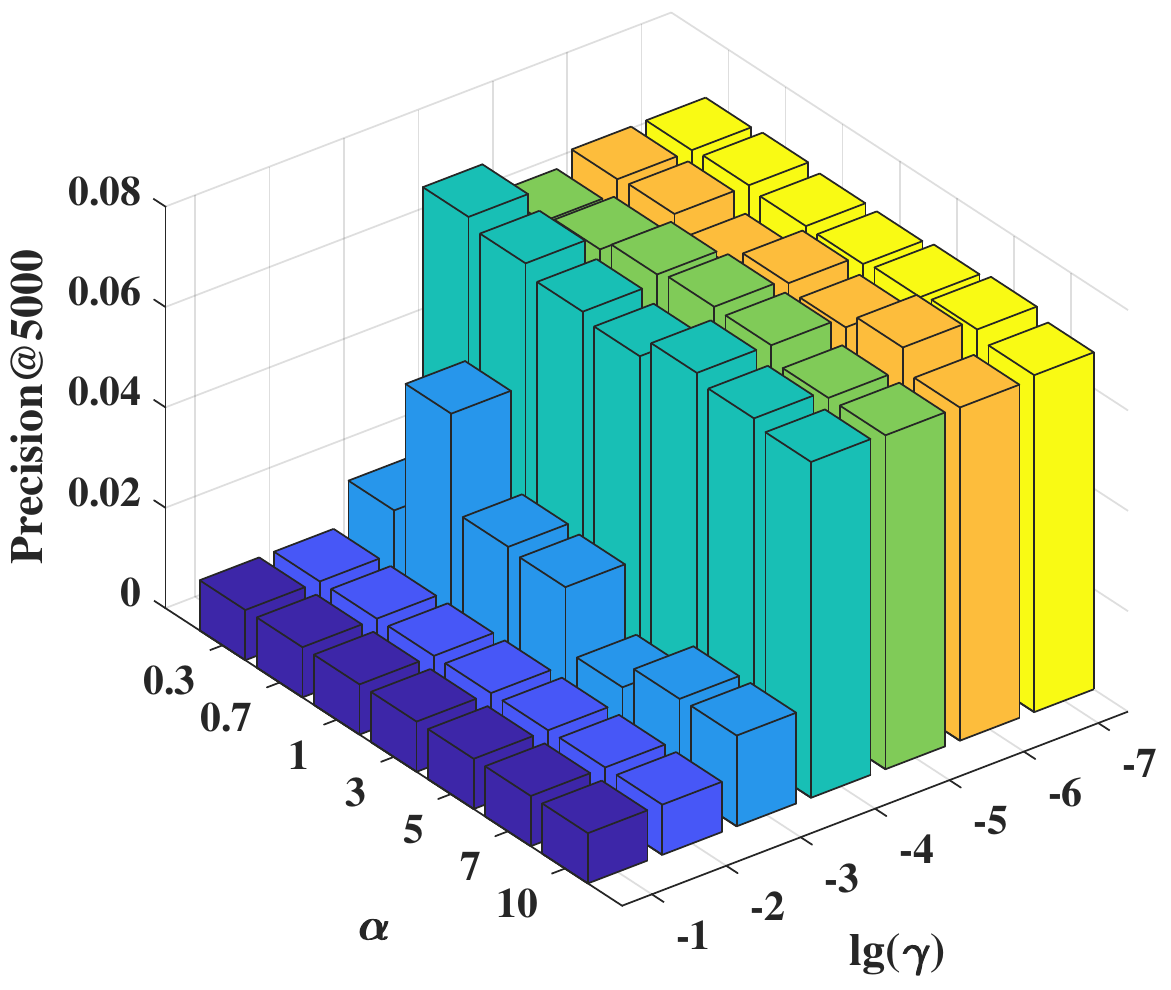}}
\caption{Subgraphs (a)-(c) show the performance in terms of the precision scores based on three  benchmark datasets. Subgraphs (d)-(f) show  the precision scores with different settings of $\alpha$ and $\gamma $, for three benchmark datasets when the hash code length is 10.}
\end{figure}

\subsection{Experimental Results and Analysis}
We compared the proposed RSLH with following methods:
Spectral Hashing (SH) \cite{weiss2009spectral},
Principle Component Analysis (PCA)-Iterative Quantization (PCA-ITQ)  \cite{gong2011iterative},
PCA-Random Rotation  (PCA-RR) \cite{gong2011iterative},
Collective Matrix Factorization Hashing (MFH) \cite{Ding2014Collective},
Supervised Discrete Hashing (SDH) \cite{shen2015supervised},
Natural Supervised  Hashing (NSH) \cite{liu2016natural},
Fast Scalable Supervised  Hashing (FSSH) \cite{luo2018fast},
Supervised Short-Length Hashing (SSLH) \cite{liu2019SSLH},
and Supervised Discrete Hashing With Mutual Linear Regression (SDHMLR) \cite{liu2019SDHMLR}.
SH, PCA-ITQ, PCA-RR, and MFH are unsupervised hashing methods, while all other hashing methods are supervised. Furthermore, all hyperparameters of these baselines were initialized as suggested in the original publications. We performed five runs for above baselines and the proposed method, and then averaged the performances for comparison. However, only nondeep methods were considered for comparison because the proposed method was linear-model-based.

\begin{table*}[htp]
  \centering
  \fontsize{9}{11}\selectfont
  \begin{threeparttable}
  
  \begin{tabular}{c|c|c|c|c|c|c|c|c|c|c|c|c}
    \toprule
     \multirow {2}{*}{Method} &\multicolumn{4}{|c}{CALTECH-101}  &\multicolumn{4}{|c}{CIFAR-10} &\multicolumn{4}{|c}{ImageNet-100}\cr
    \cmidrule(lr){2-5} \cmidrule(lr){6-9}\cmidrule(lr){10-13} 
     &\!8 bits\!&\!10 bits\!&\!12 bits\!&\!14 bits  \!&4 bits\!&6 bits\!&\!8 bits\!&\!10 bits \!&8 bits\!&10 bits\!&\!12 bits\!&\!14 bits \cr
    \midrule
    S&0.4060&0.4446&0.4689&0.4914&0.4820&0.5610&0.6135&0.6250&  0.1313&0.1740&0.2009&0.2300\cr
     G& 0.4015&0.4468&0.4698&0.4895&0.4445&0.5382&0.6104&0.6283&0.1292&0.1735&0.2068&0.2293\cr
    S+OB  &0.4079 & 0.4483  &0.4776 & 0.4842 & 0.4747  & 0.5962 & 0.6106 & 0.6474 & 0.1308 & 0.1758 & 0.2037 & 0.2270 \cr 
    S+MB&0.4227&0.4426  & 0.4811& 0.4833 &  0.4701  &  0.5964& 0.6278 & 0.6356&0.1304 & 0.1807 & 0.2074 &0.2282\cr 
    G+OB&0.4235&0.4504 &  0.4670 & 0.4846 & 0.3819   & 0.5919 &0.6135 & 0.6438 &0.1315 &0.1765 & 0.2072 & 0.2332\cr
      G+MB& 0.4263&0.4469&0.4712&0.4837& 0.4688  &  0.5934 & 0.6163  &  0.6473&0.1358 & 0.1749 & 0.2089 & 0.2330 \cr 
    \bottomrule
    \end{tabular}
    \caption{Ablation study in terms of mAP score on three benchmark datasets. }
   \end{threeparttable}
\end{table*}

\begin{figure}[ht]
\centering
\subfigure[SSLH]{
\includegraphics[width=0.23\textwidth]{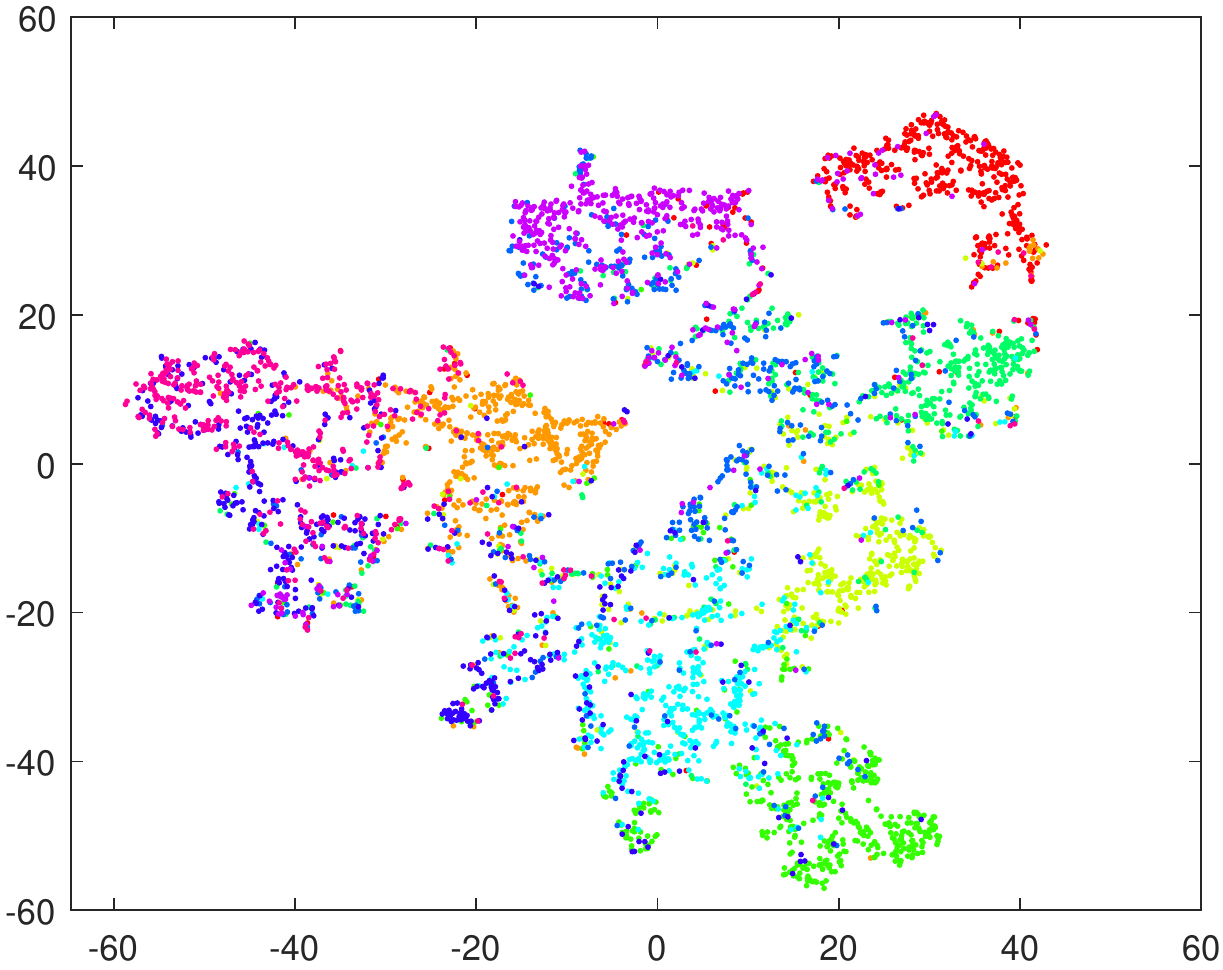}}
\subfigure[RSLH-B]{
\includegraphics[width=0.23\textwidth]{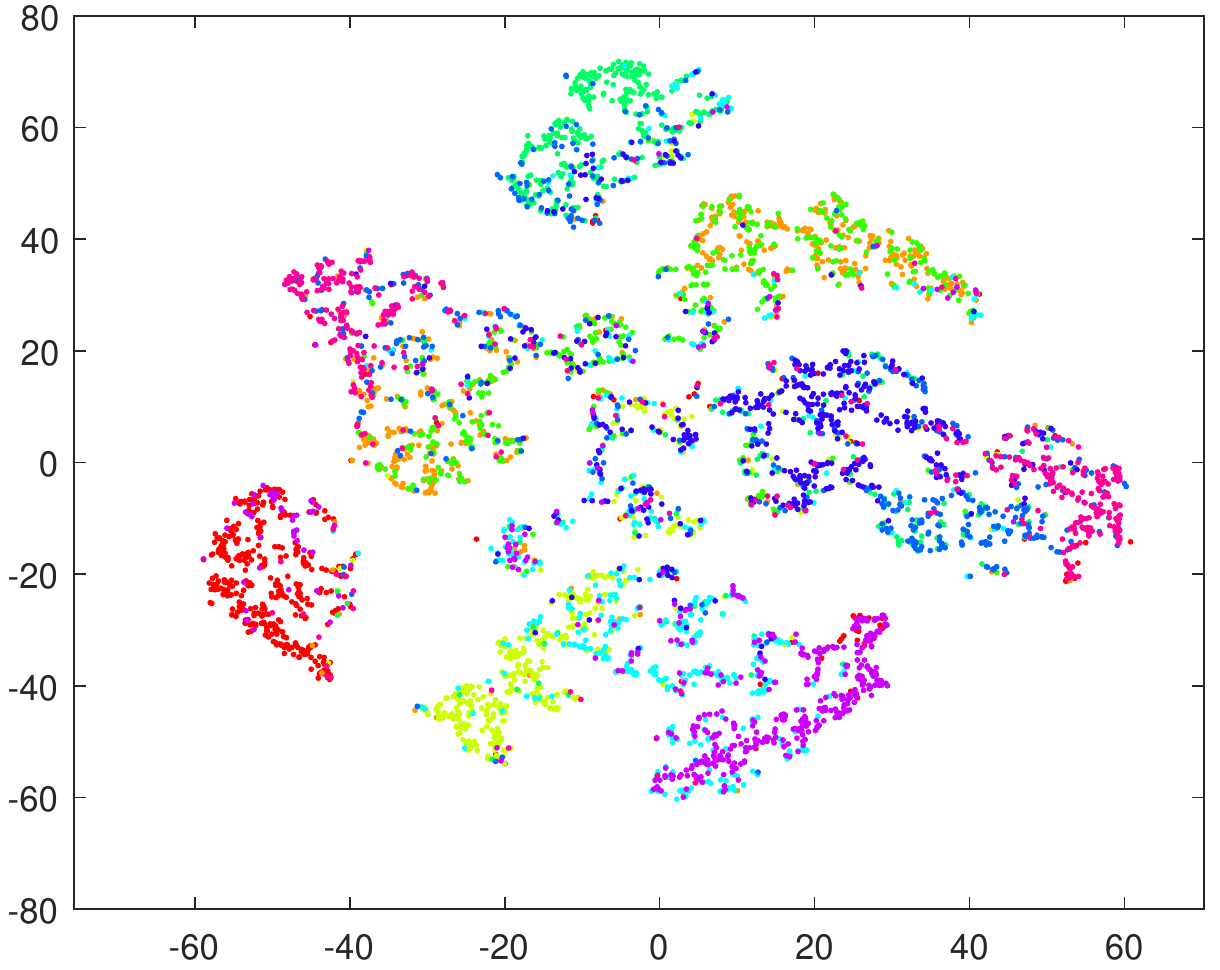}}
\subfigure[RSLH]{
\includegraphics[width=0.23\textwidth]{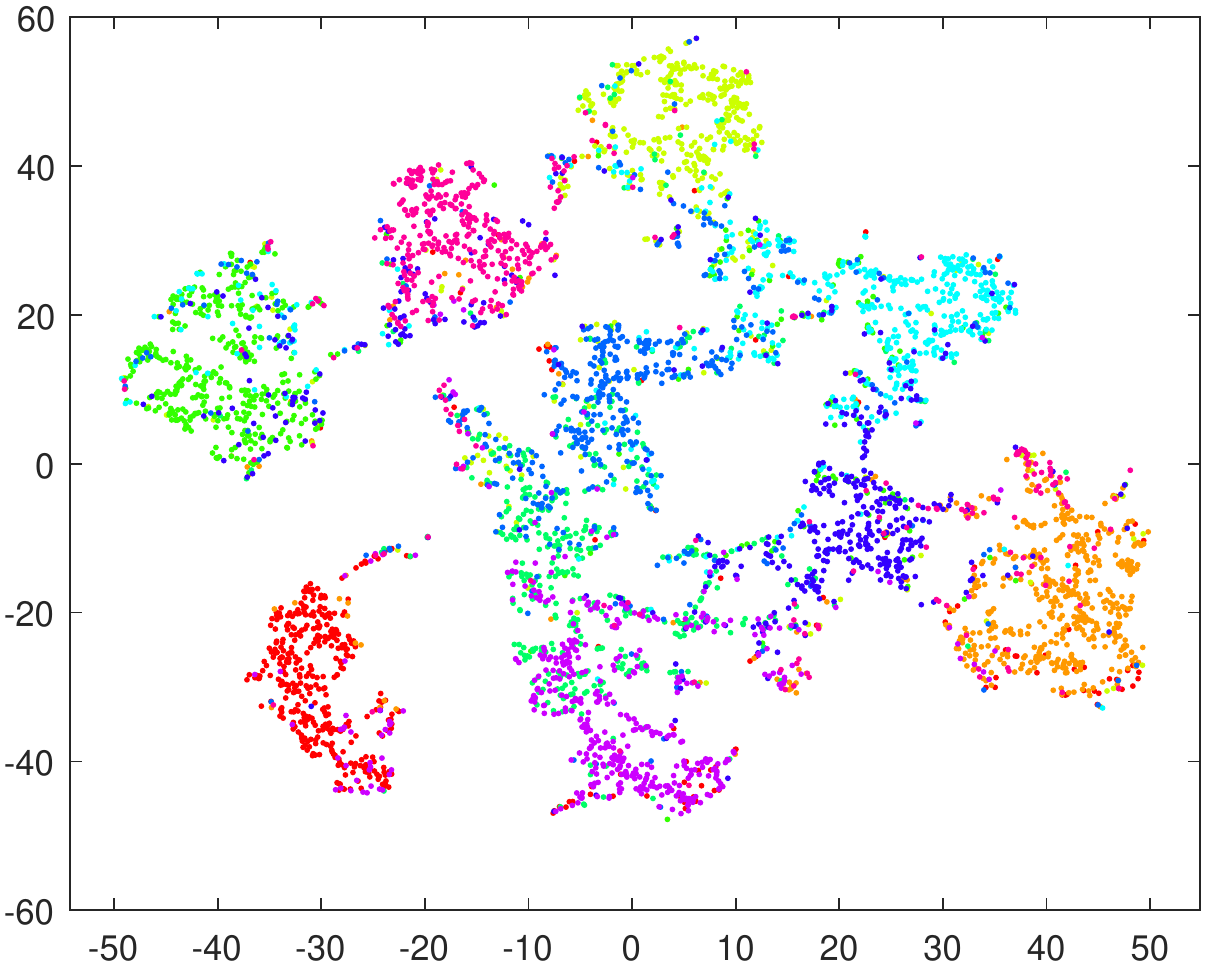}}
\subfigure[Convergence]{
\includegraphics[width=0.23\textwidth]{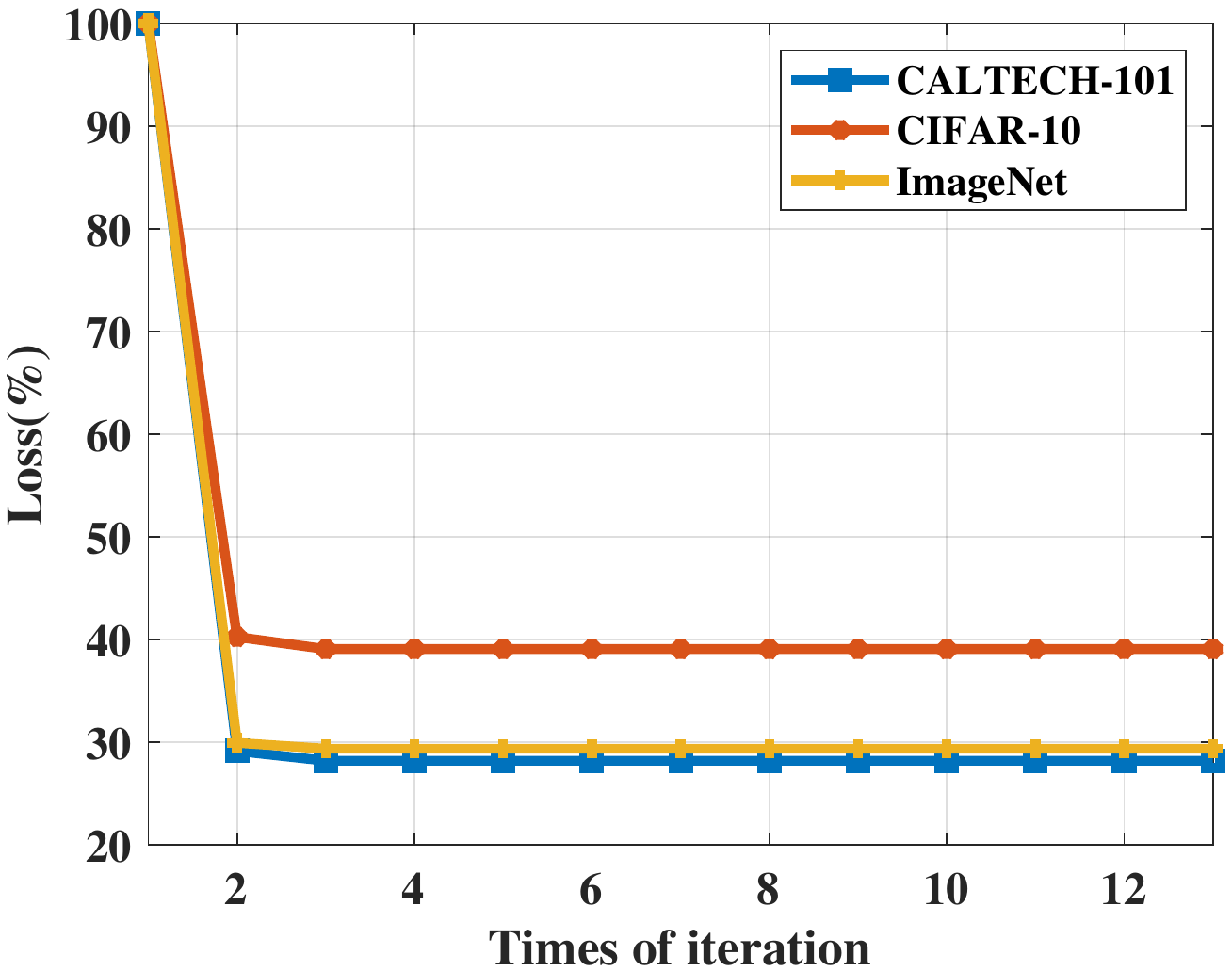}}
\caption{Subgraphs (a)-(c) show the t-SNE visualization of hash codes on CIFAR-10. The length of hash code is 4. Subgraph (d) shows the convergence curves of the proposed RSLH  for three datasets. In each curve, the loss of the proposed RSLH for the first iteration is considered to be 100\%. The length of hash code is 10. }
\end{figure}

In the experiments, the short length, $L$, is slightly greater than the $log_{2}(c)$ value that are approximately 6.7, 3.3 and 6.7 in the three datasets CALTECH-101, CIFAR-10 and ImageNet-100, respectively. We approximately set the short-length as not greater than 10. 

The top panel of Table 1 lists the mAP values for each method, for three datasets, CALTECH-101, CIFAR-10 and ImageNet-100. The mAP performance of the RSLH is considerably better than those of the other methods for these three benchmark datasets, with short-length hash codes. 
Specifically, compared to the best unsupervised hashing methods, we achieved absolute boosts of 22.09\%, 30.65\% and 15.16\% in average mAP for different bits on CALTECH-101, CIFAR-10 and ImgageNet-100, respectively. Compared to the state-of-the-art  short-length hashing methods SSLH,  we obtained absolute boosts of 4.26\%, 4.31\% and 4.19\% in average mAP for different bits on the three datasets, respectively.

The bottom panel of Table 1 depicts the mAP@H$\leq$2 value for each method, for three datasets, CALTECH-101, CIFAR-10 and ImageNet-100. The  mAP@H$\leq$2 performance of the RSLH is obviously better than those of the other methods for these three benchmark datasets, with short-length hash codes. Specifically, compared to the best unsupervised hashing methods, we achieved absolute boosts of  30.31\%, 29.81\% and 13.85\% in average mAP@H$\leq$2 for different bits on CALTECH-101, CIFAR-10 and ImgageNet-100, respectively. Compared to the state-of-the-art  short-length hashing methods SSLH,  we obtained absolute boosts of 7.79\%, 4.17\% and 2.89\% in average mAP@H$\leq$2 for different bits on the three datasets, respectively.

Substantial improvement can also be seen in Figure 1 suggraphs (a)-(c), in terms of the precision score, where the comparison between the proposed RSLH and the existing methods is depicted for different lengths of hash codes. In this study, RSLH-B indicates the proposed method without model boosting. The RSLH will exhibit considerably better performance if the length of hash code length is shorter. However, the improvement will reduce when the length is larger than 32 bits, indicating that the proposed RSLH has a distinct advantage with short-length hash codes

Table 2 shows the ablation study in terms of mAP score on three benchmark datasets, where $\bf{S}$, $\bf{G}$, $\bf{MB}$ and $\bf{OB}$ indicate the proposed method using the original similarity matrix $\bf{S}$, using the term $\bf{G}$ instead of $\bf{S}$ in Eq. (13), original model boosting in \cite{liu2019moboost} and the proposed modified model boosting, respectively. It can bee seen that the proposed method using modified boosting strategy achieved superior performance. Moreover, the utilization of $\bf{G}$ in the proposed method does not bring  much attenuation in performance compared with original pairwise similarity matrix. However, the space complexity is reduced due to the low dimension of $\bf{G}$.

In order to verify the parameter sensitivity of the proposed method, we conducted experiments with different parameter settings. Due to limited space, we only showed the results about ${\alpha}$ and ${\gamma}$, which are most relevant to the performance of the proposed method. Figure 1 suggraphs (d)-(f)  shows the precision score of the RSLH, when ${\alpha}$ and ${\gamma}$ are within a range; the RSLH method exhibits acceptable stability and sensitivity with short-length hash codes.

In Figure 2, subgraphs (a)-(c) show the t-SNE visualization \cite{maaten2008visualizing} of the hash codes learned by  the best short-length hashing baseline SSLH and the proposed RSLH  on CIFAR-10 dataset. We can observe that the hash codes generated by RSLH show  more clear diacritical structures than SSLH, of which the hash codes in various categories are not well separated. This verifies that the hash codes generated by RSLH are more diacritical than those of SSLH, enabling more effective image retrieval. In addition, the proposed model boosting shows a little more clear structures than the original method, verifying its efficiency.
Subgraphs (d) depicts the changes in the objective values achieved by the RSLH for three datasets. As the number of iterations increases, the objective values become small and stable, indicating that the RSLH converges rapidly during training, thereby distinctly reducing the time required for training.

\section{Conclusion}
In this study, we propose a method for short-length hashing, wherein the semantic label information is leveraged by mutual regression and  asymmetric pairwise similarity preserving, while the uncorrelation constraint is approximated by orthogonal representation. In addition, the balance constraint is achieved based on a modified model boosting framework. Extensive experiments conducted on three image benchmarks indicate  superior performance of the proposed method, compared to the other existing methods. In future, we will attempt to accelerate the training  time and extend the proposed framework to nonlinear-based models.


\end{document}